\documentclass[12pt,a4paper]{article}

\newcommand{\la}{\lambda}
\newcommand{\de}{\delta}
\newcommand{\hm}{\hat{\mu}}
\newcommand{\hn}{\hat{\nu}}
\newcommand{\ga}{\Gamma}
\newcommand{\ba}{\mbox{\boldmath$A$}}
\newcommand{\bff}{\mbox{\boldmath$F$}}
\newcommand{\bga}{\mbox{\boldmath$\Gamma$}}
\newcommand{\I}{\mathrm{I}}

\begin{document}

\begin{flushright}
hep-th/0010007
\end{flushright}
\vspace{1.8cm}

\begin{center}
\textbf{\Large Gauge Symmetry Breaking and \\
Generalized Monopole in Non-BPS D-Brane Action}
\end{center}
\vspace{1.6cm}
\begin{center}
Shijong Ryang
\end{center}

\begin{center}
\textit{Department of Physics \\ Kyoto Prefectural University of Medicine
\\ Taishogun, Kyoto 603-8334 Japan}
\par
\texttt{ryang@koto.kpu-m.ac.jp}
\end{center}
\vspace{2.8cm}
\begin{abstract}
Based on the DBI action for the four coincident non-BPS D9-branes in the
type IIA string theory we demonstrate that the gauge symmetry breaking
through the tachyon condensation into the generalized monopole of 
codimension five produces a pair of two coincident BPS D4-branes. The
nontrivial gauge field configuration is studied and shown to yield
the non-zero generalized magnetic charge. We discuss how this explicit
demonstration is related with the higher K-theory group.
\end{abstract}
\vspace{3cm}
\begin{flushleft}
October, 2000
\end{flushleft}
\newpage
\section{Introduction}

In the study of non-perturbative behaviors of the supersymmetric 
string and gauge field theories the D-branes have offered various 
important aspects. The spectrum in some vacua of the string theory
includes not only the BPS D-branes that are themselves solitonic
stable states, but also the unstable non-BPS D-branes that 
can be made stable by modding out with some finite
group \cite{AS,BG,SLS}. Besides the unstable non-BPS D-branes there 
is an unstable object that consists of an equal number of BPS D$p$-branes
and BPS D$p$-antibranes (D$\bar{p}$-branes), whose instability is due 
to the presence of a tachyon in the spectrum of the $p-\bar{p}$ open 
strings. In the type IIB string theory the stable D-brane states of
lower dimensions appear as bound states of the D$p$-brane  
D$\bar{p}$-brane system via the tachyon condensation and the D-brane 
charges are identified with the elements of $\tilde{\mathrm{K}}(X)$, 
the reduced K-theory group of the spacetime 
manifold X  \cite{EW}. In this approach 
from a D9-D$\bar{9}$ pair and two D9-D$\bar{9}$ pairs via the tachyon
condensations a BPS D7-brane and a BPS D5-brane are built respectively
where the nontrivial gauge field configurations are given by the 
two-dimensional vortex and the four-dimensional instanton, which are 
associated with the homotopy groups and the reduced K-theory groups
in such ways as $\Pi_{1}(U(1)) = \tilde{\mathrm{K}}(S^2) = 
Z$ and $\Pi_{3}(U(2)) = \tilde{\mathrm{K}}(S^4) = Z$. 
For the type IIA string theory a non-BPS D9-brane
itself is so unstable that a number of unstable D9-branes dcay to produce
the BPS D-branes by the tachyon condensations, 
which are classified by the higher K-theory group 
$\mathrm{K}^{-1}(X)$  \cite{PH}. For instance a system of 
two D9-branes and a system of the four D9-branes 
decay to yield a D6-brane and a D4-brane respectively 
through the 't Hooft-Polyakov three-dimensional monopole and a 
generalized five-dimensional monopole which are classified by 
$\Pi_2(U(2)/U(1)\times U(1)) = \mathrm{K}^{-1}(S^3) = Z$ 
and $\Pi_4(U(4)/U(2) \times U(2)) = \mathrm{K}^{-1}(S^5) = Z$. 
On the other hand the unstable D9-branes
can produce the D8-D$\bar{8}$ pairs by the tachyon condensation into a 
kink not accompanied with the nontrivial gauge configuration.

The non-BPS Wess-Zumino couplings of type IIA non-BPS D-branes to the 
Ramond-Ramond (R-R) potentials have been presented and shown to produce
the BPS Wess-Zumino action through the tachyon condensation \cite{BCR}.
For the system of D-branes and D-antibranes of the type II string theories
the generalization of the Wess-Zumino action has been also performed
\cite{KW}. Moreover the effective DBI actions for non-BPS D-branes
in the type II string theories have been proposed by Sen \cite{ASE}
where the interactions between the tachyon and other light fields are
restricted by the supersymmetry and the general covariance and the
requirement of possible tachyon condensation. As an attempt to generalize
Sen's proposal an effective non-BPS D-brane action has been presented
in Ref. \cite{JK}. Based on this effective action in the type IIA or IIB
string theory a BPS D$p$-brane is shown to be produced from a non-BPS 
D$(p+1)$-brane via the tachyon condensation in a form of kink.
Repeating this step all D-branes are constructed from a number of 
D9-branes or a system of D9-branes and D$\bar{9}$-branes. 
This step by step construction uses a kink solution where the gauge fields
are trivial. On the other hand in the K-theory group view-point for the 
BPS D-brane charges the nontrivial gauge field configurations play 
important roles. The other attempt to generalize Sen's proposal has been
performed by requiring the T-duality \cite{BRW} 
to yield the different type of the non-BPS D-brane action, whose 
expression was suggested by calculations of S-matrix elements \cite{MG}.
There is a proposal of an interpolating DBI action for a 
single non-BPS D-brane which reduces to the previous two types 
of actions in the two particular limits \cite{JKL}.

Based on the effective DBI action for the non-BPS D9-branes in the type
IIA string theory presented in Refs. \cite{BRW,MG} we will investigate 
the tachyon condensation in a form of the nontrivial gauge field 
configuration. Specially considering the Higgs mechanism in a
genralized five-dimensional monopole configuration we will construct
the world-volume gauge theories for BPS D4-branes by one step procedure.
The nontrivial gauge fields will be estimated and from them a non-zero
generalized magnetic charge will be evaluated. The explicit construction
of BPS D4-branes will be argued in comparison with the general approach
based on the higher K-theory group.

\section{One step construction}

We start to write down the DBI action for the massless bosonic excitations
of the $N$ coincident non-BPS D9-branes in the type IIA string theory
\begin{equation}
S = -C_9 \int d^{10}\sigma \mathrm{Tr}(g(T)\sqrt{-\det(G_{\mu\nu} + 
B_{\mu\nu}+ F_{\mu\nu} + D_{\mu}TD_{\nu}T )} ),
\label{act}\end{equation}
where $G_{\mu\nu}, B_{\mu\nu},  F_{\mu\nu}$ and $T$ are the metric, 
Neveu-Schwarz antisymmetric tensor, $U(N)$ gauge field strength and 
tachyon field respectively. The massless fermionic excitations can be
included by requiring the spacetime supersymmetry invariance with no 
$\kappa$-symmetry. The gauge trace is taken as a symmetrized trace
\cite{AT}. This type of abelian DBI action for a single non-BPS 
D-brane was proposed from the viewpoint of T-duality \cite{BRW}. 
Here we have taken the trace operation in the same way as was made in
Ref. \cite{MG}, which is slightly different from the trace operation
in the other type of DBI action for non-BPS D-branes in Ref. \cite{JK}
where the trace is separately taken in the potential part and the
sruare root part. The tachyon field $T$ has a tendency to role down to a
certain value $T_0$ at the minimum of the potential $g(T)$. 
We express the potential as $g(T) = \I + V(T)$  where 
I is  the $4\times 4$ unit matrix and $V(T)$ denotes universal 
tachyon potential \cite{ASN}. The potential $g(T)$ is conjectured
to vanish at $T = T_0$. When we restrict ourselves to tachyon part of the
DBI action (\ref{act}) by putting all the fields zero except the tachyon 
field, the negative energy density of the condensed tachyon is conjectured
to cancel the positive energy density of the $N$ D9-branes asymptotically,
which is expressed as $C_9 \mathrm{Tr}V(T_0) + NC_9 = 0$ 
where $C_9$ is the tension
of a single non-BPS D9-brane. 

Now we consider a particular system of four coincident non-BPS D9-branes
with flat metric and no $B_{\mu\nu}$ background. In the following we will
work in the static gauge. The expansion of the square root in (\ref{act})
yields
\begin{equation}
S = -C_9 \int d^{10}\sigma \mathrm{Tr}( \I + V(T) )( \I 
+ \frac{1}{4}F_{\mu\nu}F^{\mu\nu}  + \frac{1}{2} D_{\mu}TD^{\mu}T ),
\label{roo}\end{equation}
where $D_{\mu}T = \partial_{\mu}T + i[A_{\mu}, T], F_{\mu\nu} = 
\partial_{\mu}A_{\nu} - \partial_{\nu}A_{\mu} + i[A_{\mu}, A_{\nu}]$
and the $U(4)$ gauge field $A_{\mu}$ as well as the tachyon field $T$
are in the adjoint of $U(4)$. It is natural to suppose that $V(T)$
is even and so takes the form
\begin{equation}
V(T) = - \mu^2 T^2 + \la T^4.
\end{equation}
The parameters, $\mu$ and $\la$ will be related so that the potential
$g(T)$ vanishes at $T = T_0$. 

The higher K-theory group indicates that in an unstable system of the four
D9-branes the gauge symmetry breaking occures from $U(4)$ to 
$U(2)\times U(2)$ through the tachyon condensation, where the tachyon 
field takes values in the vacuum manifold $U(4)/U(2)\times U(2)$
far from the core of the topological magnetic defect of 
codimension 5 \cite{PH}. The generators of $U(4)$ are divided into 
those of the unbroken gauge symmetry group $U(2)\times U(2)$ and
those of the coset, while they are defined by the $U(1)$ generator
$\la_0 = \I$ and the $SU(4)$ traceless 
generators $\la_a (a = 1, \cdots, 15)$. There are convenient bases of
$su(4)$ Lie algebra
\begin{eqnarray}
(E_{ij})^{kl} &=& \de_i^k \de_j^l - \de_i^l \de_j^k,  \nonumber \\
(F_{ij})^{kl}  &=& i(\de_i^k \de_j^l + \de_i^l \de_j^k), \;
 \mbox{for $i \neq j$}, \\
(F_{ii})^{kl}  &=& i(\de_i^k \de_i^l - \de_4^l \de_4^k) \nonumber
\end{eqnarray}
with $i = 1, \cdots, 4$, which are related with $\la_a$ in three groups as
$\{i\la_1 = F_{12}, i\la_2 = E_{12}, i\la_3 = F_{11} - F_{22} \}, 
\{ i\la_4 = F_{13}, i\la_5 = E_{13}, i\la_6 = F_{23}, i\la_7 = E_{23}, 
i\la_8 = (F_{11} + F_{22} - 2F_{33})/\sqrt{3}, i\la_9 = F_{14},
i\la_{10} = E_{14}, i\la_{11} = F_{24}, i\la_{12} = E_{24} \}, 
\{i\la_{13} = F_{34}, i\la_{14} = E_{34}, i\la_{15} = (F_{11} + F_{22}
+ F_{33})/\sqrt{6} \}$. In view of these expressions the unbroken gauge
symmetry has one $U(2)$ generators $\bar{t}^a$ consisting of $F_{12}, 
E_{12}$  and $F_{11} - F_{22}$ with
$( \I - i(F_{11} + F_{22} - F_{33}))/2$ and the other $U(2)$ ones
$t'^a$ consisting of $F_{34}, E_{34}$ and $F_{33}$ with 
$( \I + i(F_{11} + F_{22} - F_{33}))/2$. The remaining eight bases such
as $E_{13}, E_{14}, E_{23}, E_{24}$ and $F_{13}, F_{14}, F_{23}, F_{24}$
yield the generators $\tilde{t}^{\alpha}$ of the coset. Here we express
the $U(4)$ gauge field as
$A_{\mu} = \sum_{a=1}^{4}\bar{A}_{\mu}^{a}\bar{t}^a + \sum_{a=1}^{4}
{A'}_{\mu}^{a}{t'}^a +  \sum_{\alpha=1}^{8}\tilde{A}_{\mu}^{\alpha}
\tilde{t}^{\alpha}$.

Taking the low energy limits that all derivatives and all field strength
of massless fields are put small we further expand the effective action
(\ref{roo}) as 
\begin{equation}
S = -C_9 \int d^{10}\sigma \mathrm{Tr}( \I + 
\frac{1}{4} F_{\mu\nu}F^{\mu\nu} + \frac{1}{2} D_{\mu}TD^{\mu}T + V(T) ),
\label{eff}\end{equation}
where the non-leading terms $V(T)(F_{\mu\nu}^2/4 + (D_{\mu}T)^2/2)$
have been suppressed. From the leading terms in (\ref{eff}) the equations
of motion for the tachyon field and the gauge field are given by
\begin{eqnarray}
D_{\hm}D^{\hm}T + D_iD^iT &=& -2\mu^2T + 4\la T^3, \label{eqt}\\
\partial^{\hm}F_{\hm\hn} + i[A^{\hm}, F_{\hm\hn}] + \partial^{i}F_{i\hn}
+ i[A^{i}, F_{i\hn}] &=& i[T, D_{\hn}T], \label{eqn}\\
\partial^{\hm}F_{\hm j} + i[A^{\hm}, F_{\hm j}] + \partial^{i}F_{ij}
+ i[A^{i}, F_{ij}] &=& i[T, D_{j}T], 
\label{eqj}\end{eqnarray}
where we have split the world-volume coordinates $\sigma^{\mu}$
into $(x^{\hm}, x^i)$, $\hm = 0, \cdots, 4$ and $i = 5, \cdots, 10$.
Now we consider the tachyon condensation into the topological 
defect in codimension 5 so that we can assume that the tachyon condensate
is a function of  the five transverse coordinates $x^i, i = 5,
\cdots, 10$. Therefore from (\ref{eqn}) we have a trivial classical 
solution $A_{\hm}^c = 0$ and a condition that $A_i^c$ should be 
independent of $x_{\hm}$. Then the first term of the left-handed side
(LHS) in (\ref{eqt}) and the first two terms of the LHS in (\ref{eqj})
vanish so that the remaining terms in (\ref{eqt}) and (\ref{eqj}) 
combine to yield the nontrivial generalized monopole solution
$(A_i^c(x_j), T^c(x_j))$. Far from the core the tachyon takes the form
$T_0^c(x_j) = c_0U$ with $U^2=\I$ and a constant $c_0$, whose behavior
will be discussed later. We choose a parametrization $4\la = \mu^4$
so that the potential $\mathrm{Tr}(\I 
+ V(T))$ in (\ref{eff}) vanishes at the minimum of the potential
and is expressed as $\mathrm{Tr}\la(T^2 - \mu^2\I/2\la)^2$. 
Therefore $c_0^2$ is fixed by $2/\mu^2$.
We assume that in the whole region the tachyon solution is expressed as
$T^c = c(r)U$ with $r = \sqrt{x_i^2}$, the radius in the five 
transverse  dimensions, where a real function $c(r)$ approaches  
$\sqrt{2}/\mu$ for $r \rightarrow \infty$ and becomes zero near 
$r = 0$ \cite{EW, PH}. Though the function $c(r)$ may be numerically
determined by solving (\ref{eqt}) and (\ref{eqj}), its explicit form 
is not so important for the following arguments. 

In order to consider the behaviors of 
fluctuations around the classical configuration, we put
\begin{eqnarray}
\bar{A}_{\hm} &=& \bar{A}_{\hm}^c + \bar{\ba}_{\hm}, \; {A'}_{\hm} = 
{A'}_{\hm}^c + {\ba'}_{\hm}, \; \tilde{A}_{\hm} = \tilde{A}_{\hm}^c +
 \tilde{\ba}_{\hm} , \nonumber \\   \bar{A}_{i} &=& \bar{A}_{i}^c(x_j)
+ \bar{X}_i,\;  {A'}_{i} = {A'}_{i}^c(x_j) 
+ X'_i, \; \tilde{A}_{i} = \tilde{A}_{i}^c(x_j) + \tilde{\ba}_{i} 
\label{exp}\end{eqnarray}
with $\bar{A}_{\hm}^c =  {A'}_{\hm}^c =  \tilde{A}_{\hm}^c = 0$ and
$T = T^c(x_j) + \eta$. Through the Higgs mechanism the substitutions
of the expansions (\ref{exp}) into (\ref{eff}) produce the spectrum 
consisting of the massive Higgs boson $\eta$, the massive bosons
$\tilde{\ba}_{\hm}$ as well as $\tilde{\ba}_{i}$ 
associated with the broken gauge symmetry, and the
massless bosons $\bar{\ba}_{\hm}, {\ba'}_{\hm}$
as well as $\bar{X}_i, X'_i$ associated with the 
unbroken gauge symmetry $U(2)\times U(2)$. Even though the spontaneous
symmetry breaking of gauge group $U(4)$ occurs in the $i$-components, this
must hold for the other $\hm$-components as well, since the gauge symmetry
breaking is independent of the component of particular gauge fields. 
Therefore the gauge  boson $\tilde{\ba}_{\hm}$  becomes massive in the 
same way as $\tilde{\ba}_{i}$. Here we assume that the massless 
fields $\bar{\ba}_{\hm}, {\ba'}_{\hm}, \bar{X}_i, X'_i$ have no 
$x_i$ dependence. The kinetic energy part of the gauge
field Tr$F_{\mu\nu}^2$ is decomposed into Tr$(F^2_{\hm\hn} + 2F^2_{i\hm}
+ F^2_{ij})$. In Tr$F^2_{ij}$ the expansions (\ref{exp}) yield the 
classical part and the quadratic fluctuation part which includes
the following term
\begin{eqnarray}
\mathrm{Tr}{\bff}_{ij}^2 &=& \mathrm{Tr}( \; \bar{\bff}_{ij}^2 + 
{\bff'}_{ij}^2 + \mbox{\boldmath$W$}_{ij}^2 + 2ig( [\tilde{\ba}_{i},
\tilde{\ba}_{j}]( \bar{\bff}^{ij} + {\bff'}^{ij} ) +
 \mbox{\boldmath$W$}_{ij}\ba^{ij} ) \nonumber \\
&-& g^2( [\tilde{\ba}_{i}, \tilde{\ba}_{j}]^2 + \ba_{ij}^2 )\; ),
\label{fsq}\end{eqnarray}
where $\bar{\bff}_{ij} = \partial_i\bar{X}_j -  \partial_j\bar{X}_i
+ i[\bar{X}_i, \bar{X}_j], {\bff'}_{ij} = \partial_i{X'}_j -  
\partial_j{X'}_i + i[{X'}_i, {X'}_j], \mbox{\boldmath$W$}_{ij} =
 \partial_i\tilde{\ba}_j -  \partial_j\tilde{\ba}_i, \ba_{ij} =
[\bar{X}_{i}+X'_i, \tilde{\ba}_j] + [\tilde{\ba}_i, \bar{X}_j+X'_j]$
and the $U(4)$ gauge coupling constant $g$ has been restored only 
here for convenience. The first two terms in the RHS of (\ref{fsq})
indicating the massless excitations become $-\mathrm{Tr}
([\bar{X}_i, \bar{X}_j]^2 + [X'_i, X'_j]^2 )$ 
owing to $\partial_i\bar{X}_j = \partial_iX'_j = 0$. 
Similarly the massless part of quadratic 
fluctuations in $\mathrm{Tr}F^2_{i\hm}$
is extracted as Tr$( (\bar{D}_{\hm}\bar{X}_i)^2 +
(D'_{\hm}X'_i)^2 )$ where $\bar{D}_{\hm}\bar{X}_i 
= \partial_{\hm}\bar{X}_i + i[\bar{\ba}_{\hm}, \bar{X}_i]$ 
and $D'_{\hm}X'_i = \partial_{\hm}X'_i + 
i[{\ba'}_{\hm}, X'_i]$. The $\mathrm{Tr}F_{\hm\hn}^2 = 
\mathrm{Tr}{\bff}_{\hm\hn}^2$ itself is also expressed
in the same way as (\ref{fsq}). The Higgs mechanism induces masses
for $\tilde{\ba}_{\hm}, \tilde{\ba}_{i}$ and $\eta$,
thereby removing them from the low-energy spectrum. Recombining 
the leading terms (\ref{eff}) which determine the classical configuration,
with the non-leading terms we obtain the effective low-energy action
for the massless fluctuating fields 
\begin{equation}
S = -C_9 \int d^{10}\sigma \mathrm{Tr}( \I + V(T^c) )
\frac{1}{4}( \bar{\bff}_{\mu\nu} \bar{\bff}^{\mu\nu} + 
{\bff'}_{\mu\nu}{\bff'}^{\mu\nu} )
\label{low}\end{equation}
with $\bar{\ba}_{\mu} = (\bar{\ba}_{\hm}, \bar{X}_i)$ and 
${\ba'}_{\mu} = ({\ba'}_{\hm}, {X'}_i)$. 
Owing to the factorization into the $x^{\hm}$ and $x^i$ dependent parts
we can integrate over the transverse coordinates $x^i$, where the trace
goes through the tachyon potential part to operate on the gauge
kinetic part. Therefore we derive 
\begin{eqnarray}
S &=& -C_9a \int d^{5}\sigma \mathrm{Tr}\frac{1}{4}( \bar{\bff}_{\hm\hn}
\bar{\bff}^{\hm\hn} + {\bff'}_{\hm\hn}{\bff'}^{\hm\hn}
 \nonumber \\ &+& 2\bar{D}_{\hm}\bar{X}_i\bar{D}^{\hm}\bar{X}^i + 
2D'_{\hm}X'_iD'^{\hm}X'^i - [\bar{X}_i, \bar{X}_j][\bar{X}^i, 
\bar{X}^j] - [X'_i, X'_j][X'^i, X'^j]),
\label{efa}\end{eqnarray}
where $a = \int d^5x \la(c^2(r) - \mu^2/2\la)^2$ and there remains a 
gauge symmetry $U(2)\times U(2)$.  

The obtained action is considered to represent a world-volume gauge theory
describing a system of two coincident BPS D4-branes and the other two
coincident BPS D4-branes. The transverse fluctuations of these D4-branes
are expressed by $\bar{X}_i$ and $X'_i$. 
The factor $C_9a$  can be regarded as the tension 
of the BPS D4-brane. In this way a pair of
BPS D4-branes emerge as the quantum fluctuations around the generalized
monopole solution. The last two terms in (\ref{efa})
are characteristic of the tachyon condensation into the generalized
monopole in this one step construction, compared with the kink case 
for the other processes where such terms are absent \cite{JK}. 

\section{Magnetic charge of the generalized monopole}

The generalized  monopole configuration of codimension 5 supported by
the unstable four D9-branes is classified by the nontrivial element in
the homotopy group of the vacuum manifold $\Pi_4(U(4)/U(2)\times U(2))
= Z$, which is related to the higher K-theory group of spacetime
\cite{PH}. For this codimension $n = 5$ defect, the tachyon field $T$
is so specified  by a generator of $\Pi_4(U(4)/U(2)\times U(2))$ that
$T$ maps the sphere $S^{n-1} = S^4$ at infinity in the five transverse
dimensions to the vacuum manifold $U(4)/U(2)\times U(2)$.
The world-volume of four D9-branes supports a $U(4)$ Chan-Paton bundle,
which is identified with a spinor bundle 
$\mathcal{S}$ of the group $SO(5)$ of rotations 
in the transverse dimensions. Then the tachyon condensate
for the generalized monopole configuration is given by 
\begin{equation}
T = c(r)\ga_i\frac{x^i}{r},
\label{tac}\end{equation}
which satisfies $T^2 = c^2(r)\I$ and corresponds to the previously defined
tachyon solution $T^c = c(r)U$ with the convergence factor $c(r)$.  
Here we change the numbering for $x_i$ into $i = 1, \cdots, 5$. 
In this stable defect $\ga_i$ are the $\ga$-matrices of the group $SO(5)$
described by $4\times 4$ matrices, which map from the four-dimensional 
spinor bundle $\mathcal{S}$ back to $\mathcal{S}$. For the
$\ga$-matrices obeying the Clifford algebra $\{\ga_i, \ga_i\} = 
2\de_{ij}$ with $i = 1, \cdots, 5$ we take a convenient representation,
\begin{eqnarray}
\ga_{\hat{i}} &=& \sigma_{\hat{i}}\otimes  \sigma_1 = \left(
\begin{array}{cc}0 &  \sigma_{\hat{i}} \\  \sigma_{\hat{i}} & 0
\end{array}\right), \; \mbox{for $\hat{i} =1, 2, 3$} \nonumber \\
\ga_4 &=& - \I_2\otimes \sigma_2 =  \left( \begin{array}{cc}0 &  i\I_2
 \\  -i\I_2 & 0 \end{array}\right), \; \ga_5 = \I_2\otimes \sigma_3 =
 \left( \begin{array}{cc}\I_2 &  0 \\  0 & -\I_2\end{array} \right),
\label{rep}\end{eqnarray}
where $\sigma_{\hat{i}}$ are the $2\times 2$ Pauli 
matrices and $\I_2$ is the $2\times 2$ unit matrix. As in the analysis
of the 't Hooft-Polyakov monopole and the vortex, we must require  
a finite-energy configuration for the generalized monopole that the 
covariant gradient of the order parameter, that is, the tachyon field
falls off sufficiently rapidly at large distances
\begin{equation}
\partial_i T = -i[A_i, T].
\label{cov}\end{equation}
In (\ref{cov}) taking account of the asymptotic form of (\ref{tac})
\begin{equation}
T = c_0\ga_i\frac{x^i}{r} = \frac{c_0}{r}\left( 
\begin{array}{cc} x^5\I_2 & x^{\hat{i}}\sigma_{\hat{i}} + ix^4\I_2
\\ x^{\hat{i}}\sigma_{\hat{i}} - ix^4\I_2 & -x^5\I_2 \end{array} \right)
\label{asy}\end{equation}
we obtain a nontrivial $U(4)$ gauge field
\begin{equation}
A_i = \frac{i}{4r^2}[\ga_i, \ga_j]x^j
\label{gau}\end{equation}
with $i = 1, \cdots, 5$, whose expression is suggested by the 
't Hooft-Polyakov monopole solution of winding number one \cite{HP}.  
When we take the viewpoint of the $U(4)$ Chan-Paton bundle, the asymptotic
solution (\ref{asy}) for tachyon shows that $T/c_0$ is the 
$4\times 4$ matrix that is simultaneously hermitian and unitary,
which is an element of $U(4)$ group as well as Lie algebra.
This intersection of $U(4)$ with its Lie algebra  $u(4)$ is identified
with the Grassmannian manifold $U(4)/U(2)\times U(2)$ \cite{DFN}.
We will examine whether the asymptotic solutions (\ref{asy}), 
(\ref{gau}) satisfy the Eq. (\ref{cov}).
Substituting the solution (\ref{gau}) into (\ref{cov}) and using 
\begin{eqnarray}
[\ga_{\hat{i}}, \ga_{\hat{j}}] &=& 2i\epsilon_{\hat{i}\hat{j}\hat{k}}
\left(\begin{array}{cc} \sigma^{\hat{k}}& 0 \\  0 &\sigma^{\hat{k}}
\end{array} \right) \equiv 2i\epsilon_{\hat{i}\hat{j}\hat{k}}P^{\hat{k}},
\nonumber \\ 
\;[\ga_{\hat{i}}, \ga_{4}] &=& 2i\left(\begin{array}{cc} -\sigma_{\hat{i}}
& 0  \\  0 &\sigma_{\hat{i}} \end{array} \right)\equiv 2iQ_{\hat{i}}, \;
[\ga_{\hat{i}}, \ga_{5}] = 2i\left(\begin{array}{cc} 0 & 
-\sigma_{\hat{i}} \\ \sigma_{\hat{i}} & 0 \end{array} \right)
\equiv 2R_{\hat{i}}
\end{eqnarray}
we observe that
the right-handed side (RHS) of (\ref{cov}) for $i = \hat{i}$ becomes
\begin{equation}
\frac{c_0}{r^3}( -\epsilon_{\hat{i}\hat{j}\hat{l}}x^{\hat{j}}x_{\hat{k}}
\epsilon^{\hat{l}\hat{k}\hat{m}}\ga_{\hat{m}} +
\sum_{k=4}^{5}x^k(x_k\ga_{\hat{i}} - x_{\hat{i}}\ga_k) ),
\label{coi}\end{equation}
where we have used the following commutation relations too
\begin{eqnarray}
[P_{\hat{l}}, \ga_{\hat{k}}] &=& 2i\epsilon_{\hat{l}\hat{k}\hat{m}}
\ga^{\hat{m}},\; [P_{\hat{l}}, \ga_{4}] = [P_{\hat{l}}, \ga_{5}] = 0,
\nonumber \\ 
\;[ Q_{\hat{i}}, \ga_{\hat{k}} ] &=& 2i{\de}_{\hat{i}\hat{k}} \ga_4, \;
[Q_{\hat{i}}, \ga_4] = -2i\ga_{\hat{i}}, \; [Q_{\hat{i}}, \ga_5] 
= 0,    \label{com} \\
\;[R_{\hat{i}}, \ga_{\hat{k}}] &=& -2\de_{\hat{i}\hat{k}}\ga_5, \;
[R_{\hat{i}}, \ga_{4}] = 0, \; [R_{\hat{i}}, \ga_{5}] 
= 2\ga_{\hat{i}}. \nonumber
\end{eqnarray}
The expression (\ref{coi}) agrees with the LHS of (\ref{cov})
\begin{equation}
\partial_{\hat{i}}T = c_0( \frac{1}{r}\ga_{\hat{i}} - \frac{x_{\hat{i}}}
{r^3}(\ga_{\hat{l}}x^{\hat{l}} + \sum_{k=4}^5 \ga_k x^k ))
\end{equation}
with $r^2 = x_{\hat{i}}^2 + x_4^2 + x_5^2$. We  further use 
\begin{eqnarray}
[\ga_4, \ga_5] &=& -2i \left( \begin{array}{cc} 0 & \I_2 \\ \I_2 & 0
\end{array} \right) \equiv -2iS, \; [S, \ga_{\hat{k}}] = 0,
\nonumber \\ 
\;[S, \ga_4] &=& -2i\ga_5, \; [S, \ga_5] = 2i\ga_4
\end{eqnarray}
 to express the RHS of (\ref{cov}) for 
$i = 4$ as $c_0(x^{\hat{j}}(x_{\hat{j}}\ga_4 - x_4 \ga_{\hat{j}}) +
x^{5}(x_{5}\ga_4 - x_4 \ga_{5}))/r^3$ which matches the LHS of 
(\ref{cov}) $\partial_4 T = c_0(\ga_4/r - x_4\ga_jx^j/r^3)$.
For $i = 5$ the RHS is similarly described as 
$c_0(x^{\hat{j}}(x_{\hat{j}}\ga_5 - x_5 \ga_{\hat{j}}) +
x^{4}(x_{4}\ga_5 - x_5 \ga_{4}))/r^3$ which also equals to the LHS
$\partial_5T$. 

Here we extend to the generalized monopole configuration of the
higher codimension 7 supported by the unstable eight D9-branes,
which is also classified by $\Pi_6(U(8)/U(4)\times U(4)) = Z$.
We define the $\ga$-matrices of $SO(7)$ whose spinor bundle is 
identified with a $U(8)$ Chan-Paton bundle, 
as the $8\times 8$ matrices, 
$\bga_{\hat{i}} = \sigma_{\hat{i}}\otimes \sigma_1\otimes 
\sigma_1, \bga_4 = -\I_2\otimes 
\sigma_2\otimes \sigma_1, \bga_5 = \I_2\otimes \sigma_3\otimes \sigma_1,
\bga_6 = -\I_2\otimes \I_2\otimes \sigma_2$ and $\bga_7 = \I_2\otimes 
\I_2\otimes \sigma_3$ which satisfy the Clifford algebra
manifestly. This definition gives a natural extension of (\ref{rep})
\begin{eqnarray}
\bga_k &=& \left( \begin{array}{cc} 0 & \ga_k \\ \ga_k & 0 \end{array}
\right), \; \mbox{for $k = 1, \cdots, 5$} \nonumber \\
\bga_6 &=& \left( \begin{array}{cc} 0 & i\I \\ -i\I & 0 \end{array}
\right), \; \bga_7 = \left( \begin{array}{cc} \I & 0 \\ 0 & -\I 
\end{array} \right).
\end{eqnarray} 
Using these $\ga$-matrices and $r^2 = x_{\hat{i}}^2 + \sum_{k=4}^7x_k^2$
we express the asymptotic tachyon condensate
in the same form as (\ref{asy}) 
\begin{equation}
T = c_0\bga_{i}\frac{x^i}{r} = \frac{c_0}{r} \left( \begin{array}{cc}
x^7\I & \sum_{k=1}^{5}x^k\ga_k  + ix^6\I \\  
\sum_{k=1}^{5}x^k\ga_k -ix^6\I   & -x^7\I \end{array} \right),
\end{equation}
which is accompanied by a nontrivial gauge field in the same expression
as (\ref{gau}). For $i = \hat{i}$ the RHS of (\ref{cov}) is also
calculated by using various commutation relations as 
\begin{equation}
\frac{c_0}{r^3}( -\epsilon_{\hat{i}\hat{j}\hat{l}}x^{\hat{j}}x_{\hat{k}}
\epsilon^{\hat{l}\hat{k}\hat{m}}\bga_{\hat{m}} +
\sum_{k=4}^{7}x^k(x_k\bga_{\hat{i}} - x_{\hat{i}}\bga_k) ), 
\end{equation}
which again agrees with $\partial_{\hat{i}}T = c_0( \bga_{\hat{i}}/r - 
x_{\hat{i}}(\bga_{\hat{l}}x^{\hat{l}} + \sum_{k=4}^7 \bga_k x^k )/r^3)$.
Similarly for $i = 4, \cdots, 7$ the Eq. (\ref{cov}) is shown
to be satisfied by the generalized monopole solution of codimension 7.

Let us return to the generlized monopole solution of codimension 5.
From (\ref{gau}) the asymptotic magnetic field strength is given by
\begin{equation}
F_{ij} = \frac{i}{4r^4}((\de_i^kr^2 - 2x_ix^k )\sigma_{jk} -
 (\de_j^kr^2 - 2x_jx^k )\sigma_{ik} - \frac{x^kx^l}{4}
[\sigma_{ik}, \sigma_{jl}])
\label{fij}\end{equation}
with $\sigma_{ij} =[\ga_i, \ga_j]$. For this nontrivial gauge field
configuration we can evaluate the generalized magnetic charge 
\begin{equation}
Q_m = - \alpha \int_{S^4} \mathrm{Tr}(\frac{1}{c_0}T F \wedge F),
\label{qm}\end{equation}
whose expression is presented in Ref. \cite{BCR}. Here we have 
introduced an appropriate normalization $\alpha$. 
The integral in (\ref{qm}) is described by $\int \mathrm{Tr}
(\frac{1}{c_0}T F^{ij}F^{kl}dS^m\epsilon_{ijklm})$ 
with $dS^m\epsilon_{ijklm} = dx_i\wedge 
dx_j \wedge dx_k \wedge dx_l$, where $dS^m$ is further expressed in
terms of the unit vector $\hat{r}^m$ as $dS^m = \hat{r}^mdS$.
The Chan-Paton indices over which the trace in (\ref{qm}) is taken,
are identified with the spinor indices.
The generalized magnetic charge is determined by the
asymptotic forms of the tachyon and gauge fields. 
Substituting
(\ref{asy}) and (\ref{fij}) into (\ref{qm}) we must take the trace by 
using Tr$\ga_1\ga_2\ga_3\ga_4\ga_5 = 1$ that is given by the 
representation (\ref{rep}). There are useful commutation relations
\begin{equation}
[\sigma_{\hat{i}\hat{k}}, \sigma_{\hat{j}\hat{l}}] = 4(
-\de_{\hat{i}\hat{j}}\sigma_{\hat{k}\hat{l}} -\de_{\hat{k}\hat{l}}
\sigma_{\hat{i}\hat{j}} + \de_{\hat{i}\hat{l}} \sigma_{\hat{k}\hat{j}}
+ \de_{\hat{k}\hat{j}} \sigma_{\hat{i}\hat{l}}),
\end{equation}
which is derived from (\ref{com}). The other commutation relations are
also obtained by
\begin{eqnarray}
[\sigma_{\hat{i}4}, \sigma_{\hat{j}\hat{l}}] &=& 4(\de_{\hat{i}\hat{j}}
\sigma_{\hat{l}4} -\de_{\hat{i}\hat{l}}\sigma_{\hat{j}4}),\;
[\sigma_{\hat{i}5}, \sigma_{\hat{j}\hat{l}}] = 4(\de_{\hat{i}\hat{j}}
\sigma_{\hat{l}5} -\de_{\hat{i}\hat{l}}\sigma_{\hat{j}5}),
\nonumber \\ 
\;[\sigma_{\hat{i}4}, \sigma_{\hat{j}4}] &=& [\sigma_{\hat{i}5}, 
\sigma_{\hat{j}5}] = -4\sigma_{\hat{i}\hat{j}}, \; 
[\sigma_{\hat{i}4}, \sigma_{\hat{j}5}] = -4 \de_{\hat{i}\hat{j}}
\sigma_{45}, \;  [\sigma_{\hat{i}\hat{j}}, \sigma_{45}] = 0.
\end{eqnarray}
They are summarized into a single form
\begin{equation}
[\sigma_{ik}, \sigma_{jl}] = 4(-\de_{ij}\sigma_{kl} -\de_{kl}
\sigma_{ij} + \de_{il} \sigma_{kj}+ \de_{kj} \sigma_{il})
\end{equation}
with $i = 1, \cdots, 5$, which turn out to be the commutation relations
of the $so(5)$ Lie algebra when they are expressed in terms of $Z_{ij} =
\sigma_{ij}/4$. This expression in a single form makes it possible to
calculate the trace of (\ref{qm}) as
\begin{equation}
Q_m = -\frac{\alpha}{c_0} \int dS\hat{r}^m\epsilon_{mijkl}\mathrm{Tr}
(TF^{ij}F^{kl}) = \alpha 4!\Omega_4,
\end{equation}
where $\Omega_4 = 8\pi^2/3$ is the volume of the unit four-sphere.
This result is  compared to $\alpha2!\Omega_2$ with $\Omega_2 = 4\pi$
for the 't Hooft-Polyakov monopole charge.
Here we choose the normalization $\alpha$ as $1/4!\Omega_4$
so that the generalized monopole configuration is considered to
have unit magnetic charge.

\section{Discussion}

In order to demonstrate the tachyon condensation accompanied by the 
nontrivial gauge field configuration we have used the non-BPS
D-brane action proposed in Refs. \cite{BRW,MG}. Taking account of the 
adjoint representation Higgs mechanism in the 
tachyon condensation into the generalized 
five-dimensional monopole configuration for the world-volume gauge theory
representing the unstable system of the four non-BPS D9-branes in the
type IIA string theory, we have explicitly constructed the effective
world-volume gauge theory describing a pair of two BPS D4-branes.
We have used a representation of $SO(5)\; \ga$-matrices to extract
the nontrivial gauge field configuration from the finite-energy 
requirement and to show that the generalized 
monopole has unit magnetic charge.

At first sight, however, there seems to be a subtle difference between
our explicit demonstration based on the non-BPS DBI action and the
general framework based on the higher K-theory group in Ref. \cite{PH}.
In the latter a stable generalized monopole itself is interpreted
as a single BPS D4-brane which is produced from four coincident
non-BPS D9-branes as a bound state. On the contrary in our case a pair of
two D4-branes are produced as the quantum fluctuations around a
generalized monopole solution.

To reconcile with the argument of the higher K-theory group we 
assume the classical magnetic defect to be identified with
a classical BPS D4-brane background. The quantum gauge
fluctuations around the static magnetic solution yield a pair
of two quantum D4-branes. The magnetic charges of the 
topological defects may be identified not with the R-R charges of
the quantum D-branes but with those of the classical  D-brane
backgrounds which are classified by the higher K-theory group of
the spacetime manifold. The emergences of the lower-dimensional
BPS D-branes through the Higgs mechanism in the tachyon condensations
from the non-BPS D9-branes are reminiscent of the appearances of
the higher-dimensional fluctuating BPS D-branes in the Matrix theory
as the quantum fluctuations around the static and classical BPS
D-brane configurations with the nontrivial D-brane charges \cite{BSS}.
In the former the building block is a space-time filling non-BPS
D9-brane, whereas in the latter it is a BPS D0-brane.

A more detailed investigation of the interrelations among 
the adjoint representation Higgs mechanism in the non-BPS D-brane action,
the homotopy structure of the classical configuration space and 
the higher K-theory group is desirable to have a deeper understanding
of the features of the classical and quantum BPS D-branes in the 
dynamics of unstable D-brane systems. It is interesting
to demonstrate that our system obtained 
by the one step construction can be reproduced by the step 
by step construction and pursue whether there
are any differences between the two constructions. Another possible
extension would be to apply our prescription to the tachyon 
condensation in the system of D9-branes and D9-antibranes in the
type IIB string theory, where the instanton solution 
accompanied by the Higgs field and its generalized solutions 
may play  important roles.

\end{document}